\documentclass[%
 reprint,
 amsmath,amssymb,
 aps,
prl,
]{revtex4-2}

\usepackage{graphicx}
\usepackage{dcolumn}
\usepackage{bm}
\usepackage{todonotes}

\begin{document}

\preprint{APS/123-QED}

\title{Measuring the Distance and ZAMS Mass of Galactic Core-Collapse Supernovae \\ Using Neutrinos}

\author{Manne Segerlund}
\affiliation{%
Department of Engineering Sciences and Mathematics, Lule{\aa} University of Technology, 
SE-97187 Lule{\aa}, Sweden
\\
}%
\author{Erin O'Sullivan}%
 \email{erin.osullivan@physics.uu.se}
\affiliation{%
 Department of Physics and Astronomy, Uppsala University, Box 516, SE-75120 Uppsala, Sweden\\
}%

\author{Evan O'Connor}
\affiliation{The Oskar Klein Centre, Department of Astronomy, Stockholm
  University, AlbaNova, SE-106 91 Stockholm, Sweden}

\date{\today}

\begin{abstract}

Neutrinos from a Galactic core-collapse supernova will be measured by neutrino detectors minutes to days before an optical signal reaches Earth. We present a novel calculation showing the ability of current and near-future neutrino detectors to make fast predictions of the progenitor distance and place constraints on the zero-age main sequence mass in order to inform the observing strategy for electromagnetic follow-up.  We show that for typical Galactic supernovae, the distance can be constrained with an uncertainty of $\sim$5\% using IceCube or Hyper-K and, furthermore, the zero-age main sequence mass can be constrained for extremal values of compactness. 

\end{abstract}

\maketitle


\section{\label{sec:intro}Introduction}

The next Galactic core-collapse supernova (CCSN) will be one of the most important astrophysical events in our lifetime.  A burst of neutrinos tens of seconds in duration with individual energies $O(10 \,\mathrm{MeV})$ will be detected by neutrino experiments around the world. As neutrinos from a supernova arrive before the first light, an unprecedented multi-messenger search campaign to identify the supernova and observe the photon shock breakout will follow. However, due to potentially significant dust obscuration in the galaxy \cite{adams:2013}, the search strategy would benefit from any information about the progenitor system available from the neutrinos. Indeed, key information about the supernova is imprinted in the neutrino signal, including localization \cite{beacom:1999} and the type of remnant (black hole or neutron star) \cite{burrows:1988}. We present here a fast and novel method to determine the distance and progenitor star structure along with constraints on the zero-age main sequence (ZAMS) mass of a Galactic CCSN, which can help guide the observing strategy of electromagnetic telescopes, potentially hastening the identification of the host star as well as allowing for an estimate of the delay time between the neutrinos and photons.  

Our method builds from the procedure described in \cite{oconnor:13,horiuchi:2017}, where it was shown that neutrinos could be used to place constraints on the presupernova structure of the progenitor star.  Here, we extend and quantify the method and include predictions of intrinsic properties that are important for electromagnetic follow-up, such as the distance to the supernova and constraints, when possible, on the progenitor ZAMS mass. We improve on past work by \cite{Kachelriess:2004}, which examined how the neutronization peak imprinted in the neutrino signal can be used to determine supernova distance in a megaton water-Cherenkov detector. Our method obtains a similar sensitivity to distance as the method described in \cite{Kachelriess:2004}, but using smaller detector masses present in current and near future experiments and without relying on the separation between neutrinos and anti-neutrinos, which can take valuable time during a supernova event and adds potential sources of error. 

We demonstrate our method for the two most sensitive current neutrino experiments, as well as three near-future detectors. The two currently operational neutrino detectors considered are IceCube \cite{Aartsen:2016nxy}, a cubic-kilometer-scale neutrino detector embedded in the glacial ice at the South Pole, and Super-Kamiokande (Super-K) \cite{Fukuda:2002uc}, a 32 kton inner-volume water-Cherenkov detector located in Japan. By 2027, we expect three other large facilities to significantly contribute to this measurement: Hyper-Kamiokande (Hyper-K) \cite{HyperK:2018}, the next-generation of Super-K which will have an inner volume of 220 ktons, DUNE, a liquid argon detector in the US that will be 40 ktons \cite{DUNE:2020}, and JUNO, a 20 kton liquid-scintillator detector in China \cite{JUNO:2016}.

In order to capitalize on the early warning provided by neutrinos, most large-scale neutrino detectors are connected to the SuperNova Early Warning System (SNEWS) \cite{Antonioli:2004, kharusi:20}. The fast reporting strategy for distance and progenitor structure described here can be implemented in SNEWS to further enhance the information reported about the supernova event.

\begin{figure*}[ht]
    \centering
    \includegraphics[width=\textwidth]{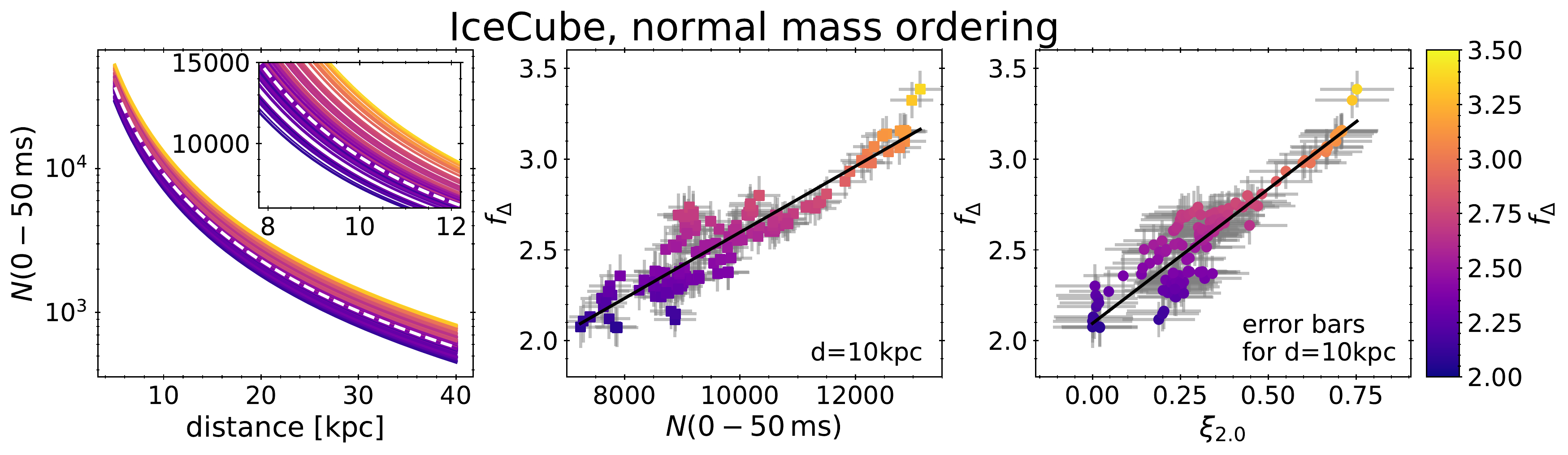}
    \caption{Progenitor dependence of the early neutrino signal in the IceCube detector assuming a normal mass ordering for the neutrinos. In the left panel we show the expected number of interactions detected by IceCube in the first 50\,ms vs. distance for 149 different progenitor models.  The color of each line denotes $f_\Delta$, a directly measurable intrinsic (although detector dependent) property of the core-collapse event.  We show this distance independent $f_\Delta$ vs. the number of counts in the first 50\,ms for a supernova at 10\,kpc (middle) and vs. the progenitor compactness (right).  The one-to-one relationships between $f_\Delta$ and these quantities allows a distance and ZAMS mass estimate from a galactic supernova event.  The 1$\sigma$ error bars shown are based on the expected Gaussian counting statistics, background level, and systematic errors from the fit to the 149 progenitor models (only for the error bar on the compactness).}
    \label{fig:IC_fdelta}
\end{figure*}

\section{\label{sec:methods}Methods}

\subsection{Tools} 

We base our analysis on the early CCSN neutrino signal generated from the evolution of 149 progenitor models from \cite{sukhbold:2016}. These models are single-star evolutions of solar-metallicity massive stars with ZAMS masses from $9.0\,M_\odot$ to $120\,M_\odot$. The presupernova structures of these models span the range expected for iron-core collapse and therefore make a complete set for this systematic study. For the core-collapse evolutions we use the FLASH \cite{fryxell:2000,couch:2013,oconnor:2018a} hydrodynamics package with an energy-dependent neutrino transport.  We use the SFHo nuclear equation of state \cite{steiner:2013} and neutrino interactions from NuLib \cite{oconnor:2015}.  In order to capture important processes which impact the neutrino signal at  early times \cite{lentz:2012}, in addition to the standard neutrino rates used in \cite{oconnor:2018a}, we utilize the microphysical electron captures rates from \cite{sullivan:2016, langanke:2003}, inelastic neutrino-electron scattering \cite{bruenn:85} and inelastic neutrino-nucleon scattering for heavy-lepton neutrinos based on \cite{thompson:2000}. Using the time evolution of the neutrino luminosity, mean energy, and the mean squared energy from our simulations, we utilize SNOwGLoBES \cite{scholberg:2012,malmenback:2019} to generate expected count rates in current and near-future neutrino detectors \footnote{We include all the neutrino data from our simulations as well as the analysis scripts for the figures and data presented in this paper in the supplemental information.}. Where stated, we use a Galactic CCSN spatial distribution from \cite{adams:2013} and a Salpeter initial mass function (IMF) \cite{salpeter:55}, i.e. $N(m)dm \propto m^{-2.35}dm$ extending from $8.75\,M_\odot$ to $130\,M_\odot$. 

\subsection{Parameter extraction methods}

The number of observed supernova neutrinos is related to distance via an inverse square law. If the early signal was progenitor-independent, then we could calculate the distance by comparing the number of observed events to the predicted signal at a known distance, a so-called standard candle approach. Indeed, this is the method utilized in \cite{Kachelriess:2004} with electron neutrinos that, during the first 10s of ms after the protoneutron star (PNS) forms, do show this behavior. However, the bulk of the early neutrino signal in many detectors consists of electron antineutrino interactions. These neutrinos do not show this universal behavior, rather the early (within the first $\sim$50\,ms) interaction rates can vary up to a factor of $\sim$2 across different progenitors. In the left panel of Figure~\ref{fig:IC_fdelta} we show the expected number of events in the first 50\,ms for our collection of 149 progenitors as a function of distance for the IceCube neutrino detector assuming the normal mass ordering and only adiabatic neutrino oscillations \footnote{Normal ordering is preferred over the inverted ordering by experiment \cite{desalas:2021}. Other detector configurations and neutrino mass orderings are available in the supplemental information}. At 10\,kpc, the mean distance to Galactic CCSN, the range of estimated counts observed in the IceCube detector in the first 50\,ms and assuming normal neutrino mass ordering  is between 7000-14000.  

To extract the distance from the detected neutrino signal we use two methods. First, we use the observed number of events detected in the first 50\,ms and compare it to an IMF-weighted average progenitor signal (the white dashed line in the left panel of Figure~\ref{fig:IC_fdelta}). This method is robust, but as mentioned above, suffers from considerable systematic uncertainty because different progenitors predict a different number of events in this time period. However, for the low-statistics regime, either at large distances or for smaller detectors, it provides the best estimate for distance. The second method relies on using the neutrino signal itself to first constrain the progenitor in a distance-independent way, then using the expected signal from that progenitor to enhance the sensitivity to distance.  As a byproduct, we obtain information about the progenitor star. The line color of the 149 individual curves (representing the 149 progenitors) in the left panel of Fig.~\ref{fig:IC_fdelta} is a parameterization of this progenitor dependence. By first constraining the progenitor, we have a more precise estimate of the expected signal. This key parameter of the early neutrino signal is

\begin{equation} 
f_\Delta = \frac{N(100-150~\mathrm{ms})}{N(0-50~\mathrm{ms})}\,, 
\end{equation}

\noindent
from Horiuchi et al. \cite{horiuchi:2017}.  $f_\Delta$ is the ratio of the number of neutrino interactions occurring between 100\,ms and 150\,ms ($N(100-150~\mathrm{ms})$) to the number of interactions occurring in the first 50\,ms ($N(0-50~\mathrm{ms})$) \footnote{In the supplemental information we show results for different definitions of $f_\Delta$}. In middle panel of Fig.~\ref{fig:IC_fdelta}, we explicitly show the key relationship we are exploiting. $f_\Delta$, which is a distance-independent quantity, has a one-to-one mapping with the expected number of interactions in the first 50\,ms.  The error bars shown in this figure are the expected 1$\sigma$ error bars for a $d=10\,$kpc supernova based on Gaussian counting statistics and also taking into account the background noise in the IceCube detector.

The combined distance estimate is achieved by averaging both of the above methods with weights corresponding to the statistical and systematic measurement error. For the statistical errors, Gaussian counting statistics is assumed with the addition for the IceCube detector of a background component \cite{Abbasi10}. The systematic error is detector specific and is based on the variance of the models to the fit (cf. the middle panel of Fig.~\ref{fig:IC_fdelta} for the IceCube detector with normal mass ordering).

Not only does $f_\Delta$ relay the expected number of events in the first 50\,ms, it is also directly related to the compactness \cite{oconnor:2011,horiuchi:2017}, a measure of the progenitor structure of the star at the end of its life. The compactness is defined as 
\begin{equation}
\xi_M = \frac{M/M_\odot}{R(M)/1000\,\mathrm{km}}\,,\label{eq:compactness}
\end{equation}
where M is some chosen mass scale (taken here to be $M=2.0\,M_\odot$ following \cite{horiuchi:2017}) and $R(M)$ is the radius that encloses that mass at the point of core collapse. This relationship between $f_\Delta$ and compactness is seen in the right panel of Fig.~\ref{fig:IC_fdelta}.  This particular relationship is distance independent, however we show the expected 1$\sigma$ error bars for a $d=10\,$kpc supernova detected in IceCube. A measurement of $f_\Delta$ allows a direct constraint on the compactness. It can be related to the ZAMS mass of the progenitor star through stellar evolution models, although the mapping is non-monotonic and can change rapidly with changing ZAMS mass \cite{sukhbold:2014}. Given the non-monotonic relationship, for a measurement of a particular value of $f_\Delta$, along with an assumption of a progenitor model series, we can determine a probability distribution for the ZAMS mass of the exploding star.

\section{\label{sec:results}Results}

\subsection{Distance}

\begin{figure}
    \centering
    \includegraphics[width=0.5\textwidth]{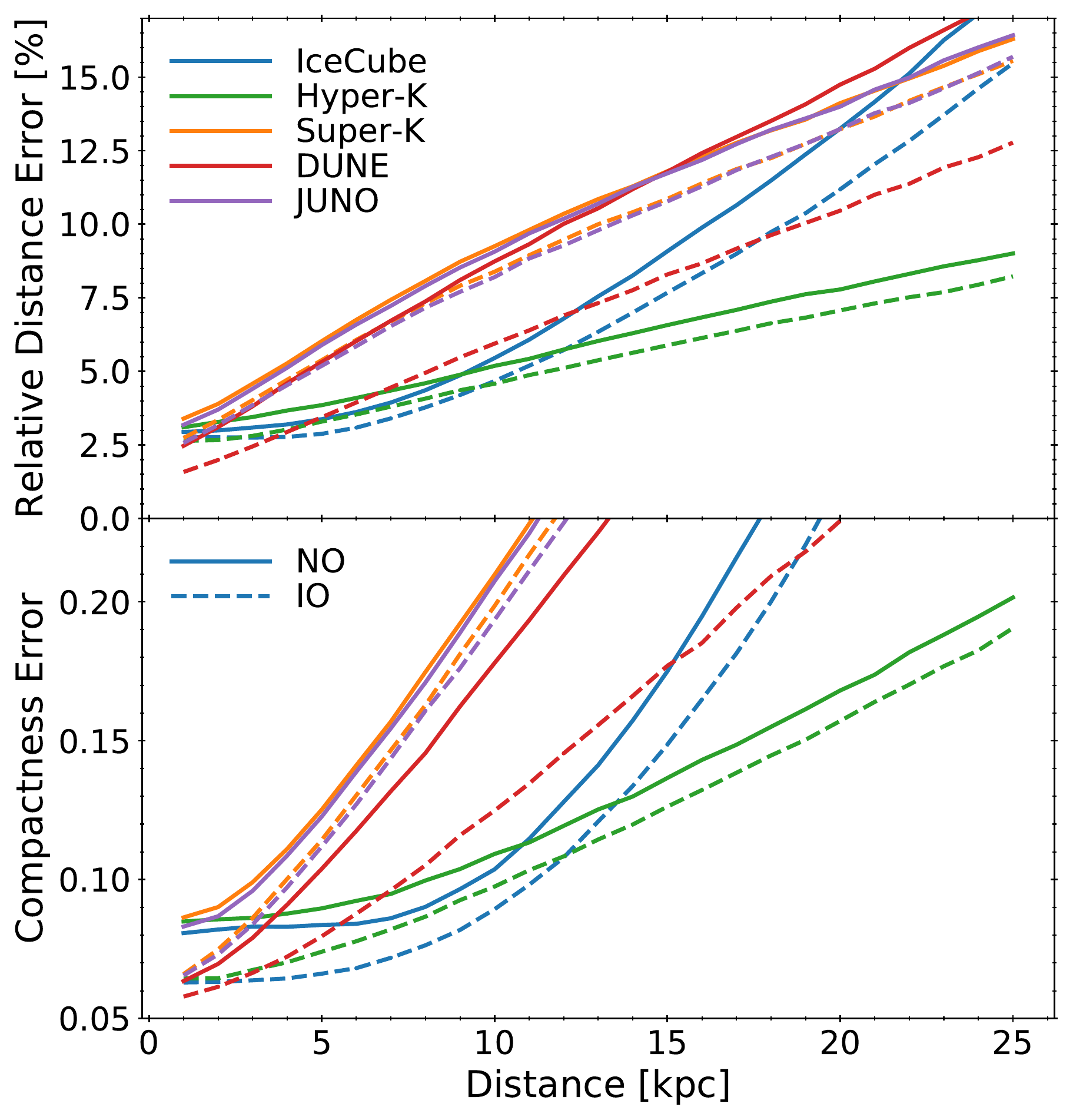}
    \caption{Estimated 1$\sigma$ errors derived from trial observations on the distance (relative; top panel) and compactness (absolute; bottom panel) marginalized over the IMF as a function of distance for each detector and the normal (NO; solid line) and inverted (IO; dashed line) neutrino mass ordering.}
    \label{fig:error_vs_distance}
\end{figure}

Following the method to determine distance outlined above, we perform a large number of mock observations to determine the precision.  For distances up to 25\,kpc in increments of 1\,kpc, for each of the five considered detectors and for each neutrino mass ordering we perform 80000 mock core-collapse events.  For each event, we randomly choose a mass based on a Salpeter IMF.  The mock observations are randomly determined based on a Gaussian distribution about the mean expected events in each window. As mentioned above, for IceCube a background noise component is added using a modified Gaussian distribution taking the spread to be $1.3\sqrt{\mu}$ where $\mu$ is the average detector background rate equal to 286\,Hz/DOM, then the mean is subtracted. The factor of 1.3 is to account for correlated hits from muons \cite{Abbasi10}. The distance is estimated for each realization and the resulting 1$\sigma$ value of the distribution of relative errors on distance ($|d-d_\mathrm{estimate}|/d$) is shown in the top panel of Fig.~\ref{fig:error_vs_distance}. 

For nearby distances (which varies detector to detector, but generally $\lesssim$ few\,kpc), the error is dominated by the systematic variation of the models from the fit shown in the middle panel of Fig.~\ref{fig:IC_fdelta}. It is worth noting that this systematic error is smallest with DUNE or in the inverted mass ordering, highlighting the fact the electron neutrinos and to some extent heavy-lepton neutrinos, are more progenitor-independent then electron antineutrinos, especially at early times. This was the original motivation for the work of \cite{Kachelriess:2004}.  As the distance increases, the errors begin to become dominated by statistics and the relative error grows linearly with distance. For IceCube, the presence of the constant background noise floor causes the error to grow faster than linear at large distances.  Marginalizing over a Galactic distance distribution \cite{adams:2013} we obtain 1$\sigma$ relative errors of 5.4\%, 8.9\%, 5.1\%, 8.3\%, and 8.7\% for IceCube, Super-K, Hyper-K, DUNE, and JUNO, respectively for the normal neutrino mass ordering \footnote{Full cumulative distributions of the relative distance error for both mass orderings and also for different choices for the definition of $f_\Delta$ are available in the supplemental information.}. Changes in the spatial distribution of CCSN events, for example, using the neutron star distribution explored in \cite{mirizzi:2006} gives similar (but $\lesssim$0.5\% larger) population-averaged relative distance errors. 

\subsection{Compactness}

In addition to extracting the distance via the early neutrino signal we can extract properties of the progenitor star itself.  Compactness, as seen in Equation \ref{eq:compactness}, is a measure of the structure of the star at the point of collapse. The original proposal from Horiuchi et al. \cite{horiuchi:2017} was to determine the compactness of the presupernova star via an observation of neutrinos.  We reproduce that analysis here, extend it to IceCube, Hyper-K, and JUNO, and quantify our ability to constrain the compactness for a galactic population. As determined by Horiuchi et al..  From the fit of $f_\Delta = m_\xi \xi_{2.0} + b_\xi$ (see right panel of Fig.~\ref{fig:IC_fdelta} for IceCube in the normal mass ordering) and an observation of $f_\Delta$, we estimate the compactness via $\tilde{\xi}_{2.0} = (f_\Delta-b_\xi)/m_\xi$, where $m_\xi$ and $b_\xi$ are the fitted slope and intercept (available in the supplemental information for each detector and neutrino mass ordering).  In the bottom panel of Fig.~\ref{fig:error_vs_distance} we show the expected 1$\sigma$ absolute error on a measurement of $\xi_{2.0}$ as a function distance. We note the same characteristics as the relative distance error.  At small distances ($\lesssim 5\,$kpc for IceCube and Hyper-K and $\lesssim 1$\,kpc for Super-K, DUNE, and JUNO) the error is dominated by systematics, it becomes linear (and statistics dominated) at larger distances. Marginalizing over a galactic distance distribution \cite{adams:2013} we obtain 1$\sigma$ absolute errors of 0.11, 0.2, 0.11, 0.17, and 0.20 for IceCube, Super-K, Hyper-K, DUNE, and JUNO respectively for the normal neutrino mass ordering \footnote{Full cumulative distributions for the absolute compactness error for both neutrino mass orderings and also for different choices for the definition of $f_\Delta$ are available in the supplemental information.}.

\subsection{Mass}

The strong correlation between $f_\Delta$ and compactness gives us an indirect measurement of the presupernova structure.  Stellar evolution--to the extent that the current modeling of the advanced burning stages, convection, and overshoot can be trusted--complicates the mapping between the ZAMS properties of the stars and the final structure at the time of core-collapse \cite{sukhbold:2014}. Furthermore, astrophysical factors, such as binarity, rotation, and metallicity will all impact the ZAMS mass to compactness mapping.  With these caveats in mind, for the single-star, solar metallicity, non-rotating model set we have chosen to use from Sukhbold et al. (2016) \cite{sukhbold:2016}, we can invert the ZAMS mass-compactness relation in order to explore potential constraints on the ZAMS mass of the progenitor star from a neutrino observation. 

\begin{figure}
    \centering
    \includegraphics[width=0.5\textwidth]{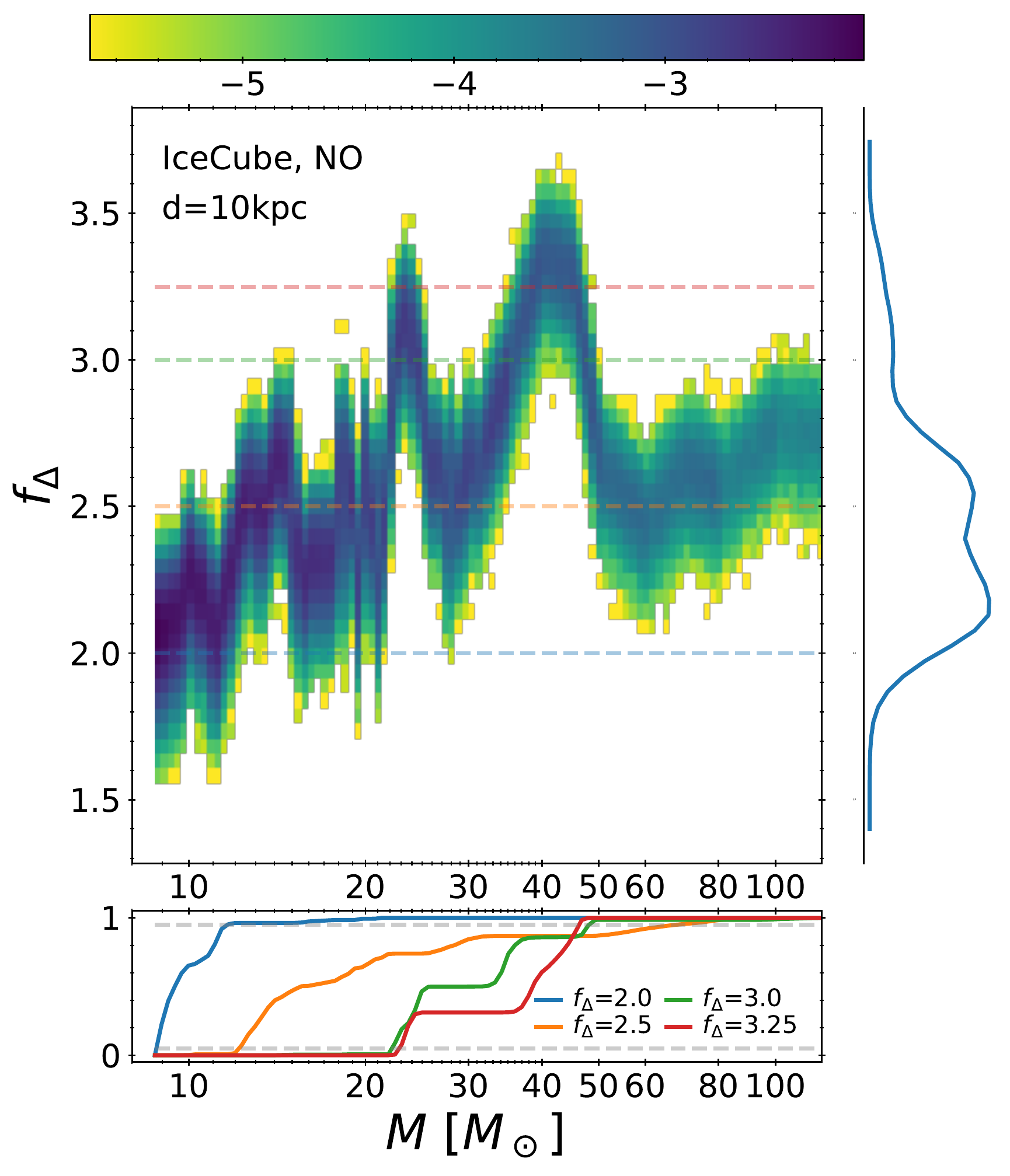}
    \caption{Probability distribution of measured $f_\Delta$ as a function of the ZAMS mass of the progenitor stars for a measured distance of 10\,kpc for the IceCube detector assuming the normal neutrino mass ordering. The color denotes the logarithm of the probability, dark blue values are $\sim$1000 times more likely than yellow.  Based on the progenitor model set we use, a measurement of $f_\Delta$ relays some information on the ZAMS mass of the progenitor star, particularly if the measured $f_\Delta$ is small. In the bottom panel we show cumulative distributions for four choices of $f_\Delta$, marked by dashed lines in the top panel.  The grey dashed lines denote cumulative probabilities of 5\% and 95\%. On the right is the marginalized distribution of observed $f_\Delta$ over the IMF.}
    \label{fig:mass_distributions}
\end{figure}

In Fig.~\ref{fig:mass_distributions}, we show the probability distribution of measured $f_\Delta$ as a function of the progenitor ZAMS mass for a supernova observed with a reconstructed distance of 10\,kpc sampled over the IMF.  We assume the IceCube detector and the normal neutrino mass ordering.  There is some structure in the $f_\Delta-M_\mathrm{ZAMS}$ plane suggesting information on the ZAMS mass may be obtained, at least in some limiting cases.  We show in the bottom panel cumulative distributions in ZAMS mass for assumed values of $f_\Delta$ = 2.0, 2.5, 3.0, and, 3.25. Note, these cumulative distributions have the statistical uncertainty of the measurement of $f_\Delta$ built in and therefore will depend on the detector and assumed distance. For an observed value of $f_\Delta$=2.0, which corresponds to progenitors of low compactness, this model set confidently places an upper ZAMS mass limit (95\% of the time) of $\sim$11.6\,$M_\odot$.  A measurement of $f_\Delta$=2.5 would give a lower ZAMS mass limit (95\% of the time) of $\sim$12.2\,$M_\odot$.  For this model set, a measurement of $f_\Delta > 3.0$ confidently ($\gtrsim$98\% of the time) places a lower limit on the ZAMS mass of 20\,$M_\odot$, and isolates potential masses to be near either $\sim 23\,M_\odot$ or 35\,$M_\odot$-50\,$M_\odot$. Even with the caveats listed above, there is general confidence in the statement that low ZAMS mass stars ($M\lesssim12\,M_\odot$) have the lowest compactness and therefore the lowest values of $f_\Delta$. It is therefore likely that for such supernovae, a constraint on ZAMS mass is possible.  

\section{\label{sec:discussion}Discussion and Conclusions}

In the results of this paper we present a compelling case that current and near-future  neutrino detectors have the capabilities and statistics to make a measurement of the distance, compactness, and a constraint on the ZAMS mass of a future Galactic CCSN.  However, both stellar evolution modelling and CCSNe evolution are complex multiphysics problems.  There are certainly systematic errors, both known and unknown, in both of these processes.  We have tried to eliminate many of the potential model dependencies.  We explore the full range of iron-core progenitors expected and show that the response we investigate is well behaved and linear over this model set.  Also, by restricting our measurements to early times, in many cases we avoid the complex multidimensional explosion dynamics and potentially avoid complex collective neutrino oscillations, that we know are present at later times. However, we note that very early explosions (prior to $\sim$150\,ms) may give smaller $f_\Delta$ then predicted here. We have tested our methods against the parameterized explosion models of \cite{warren:2020} and find no systematic differences as long as the same set of neutrino interaction rates is assumed.  On this note, we have found, though not included here, that varying these neutrino interactions (such as including and excluding neutrino-nucleon scattering and microphysical electron-capture rates) that the quantitative fits shown in Fig.~\ref{fig:IC_fdelta} can change, however the qualitative results, such as the expected precision and the ability to measure or constrain compactness and mass, are robust against these changes. We expect similar results for variations in the nuclear equation of state \cite{schneider:2019}. We therefore advocate for efforts to quantify and ideally eliminate these systematic errors so that the supernova properties discussed here can be rapidly, reliably, and accurately determined during the next Galactic CCSN and allow for optimal multi-messenger followup. These efforts would include the development and implementation of precision neutrino interaction rates and refined nuclear equations of state. Unless the core contains a large amount of angular momentum, which is the case in only a small fraction of progenitors, we do not expect rotation to affect the neutrino signal enough to distort the trends found here.

We have not included detailed detector responses, which may change the predicted number of events for a given model, or taken into account in our error estimate any uncertainties in the cross sections of neutrinos in these detectors. Furthermore, we have not combined the results from the different experiments, which may improve the distance, compactness, and mass determinations shown here.

Neutrinos from the next Galactic CCSN will provide a once-in-a-generation warning for the electromagnetic community to view the first light from shock breakout. We present a simple method to determine supernova distance and constrain the progenitor mass using the neutrino signal from current and near-future experiments, with the intention that this information could be used to aid astronomers in localizing the progenitor star, as well as inform the observing strategy. We hope this method can help to ensure we are prepared to fully maximize the data we can collect from this incredible event. 

\begin{acknowledgments}
We thank Sean Couch and MacKenzie Warren for FLASH development and access to models from \cite{warren:2020}, as well as Olga Botner, Allan Hallgren, Carlos Perez de los Heros, Christian Glaser, Kate Scholberg, and Segev BenZvi for useful discussions.  EOS and EOC would each like to acknowledge Vetenskapsr{\aa}det (the Swedish Research Council) for supporting this work under award numbers 2019-05447, 2018-04575, and 2020-00452. The simulations were performed on resources provided by the Swedish National Infrastructure for Computing (SNIC) at PDC and NSC partially funded by the Swedish Research Council through grant agreement No. 2016-07213.
\end{acknowledgments}

\providecommand{\noopsort}[1]{}\providecommand{\singleletter}[1]{#1}%

\clearpage
\appendix
\onecolumngrid

\begin{center}
\Large \textbf{Supplemental Information: Measuring the Distance and ZAMS Mass of Galactic Core-Collapse Supernovae Using Neutrinos}
\end{center}

\section{Progenitor dependence for different detectors}

\begin{figure}[h]
    \centering
    \includegraphics[width=0.8\textwidth]{IceCube_NO.pdf}\\
    \includegraphics[width=0.8\textwidth]{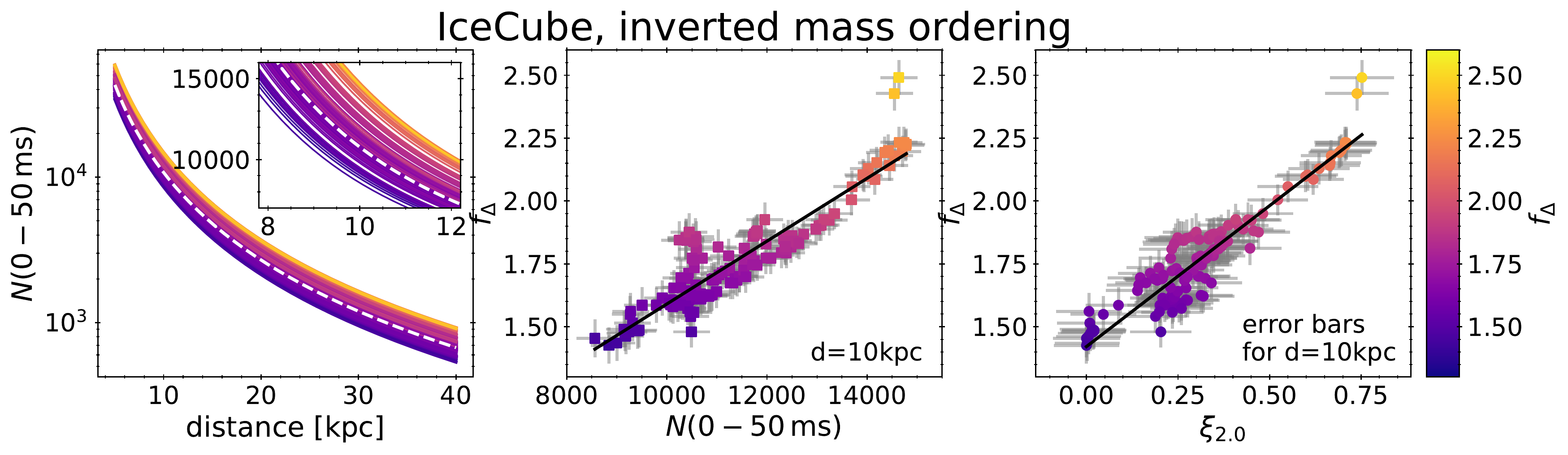}
    \caption{Progenitor dependence of early neutrino signal in the IceCube detector assuming a normal mass ordering (upper panel; repeat of Figure 1 in the main article for completeness) inverted mass ordering (lower panel) for the neutrinos. For details, see the Figure 1 caption in the main article.}
    \label{fig:IC}
\end{figure}

\begin{figure}[h]
    \centering
    \includegraphics[width=0.8\textwidth]{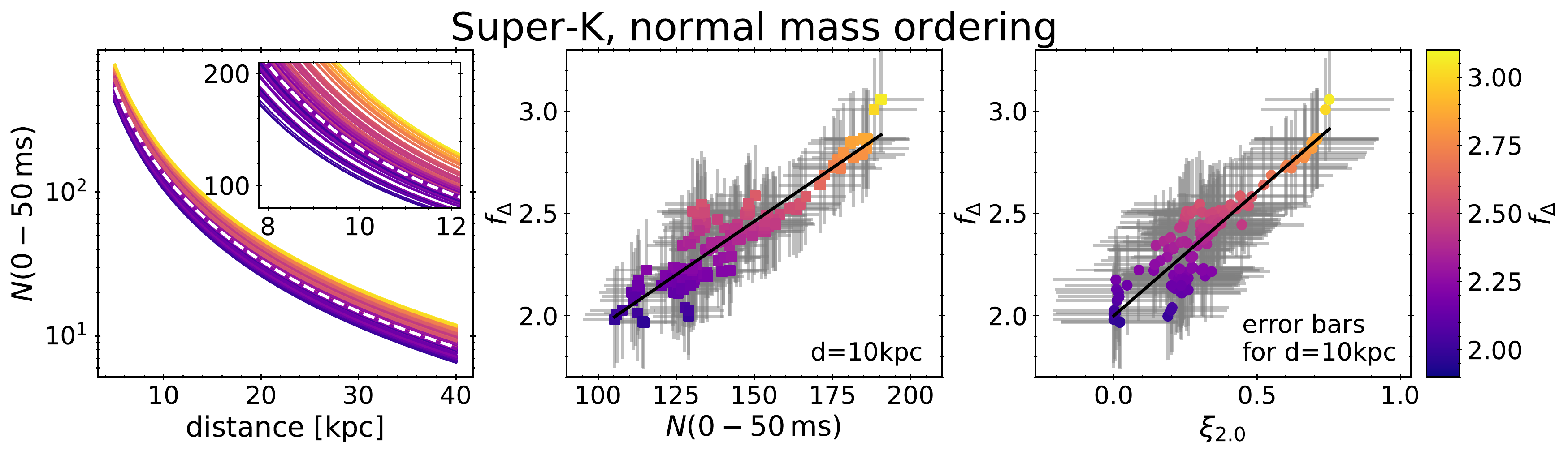}\\
    \includegraphics[width=0.8\textwidth]{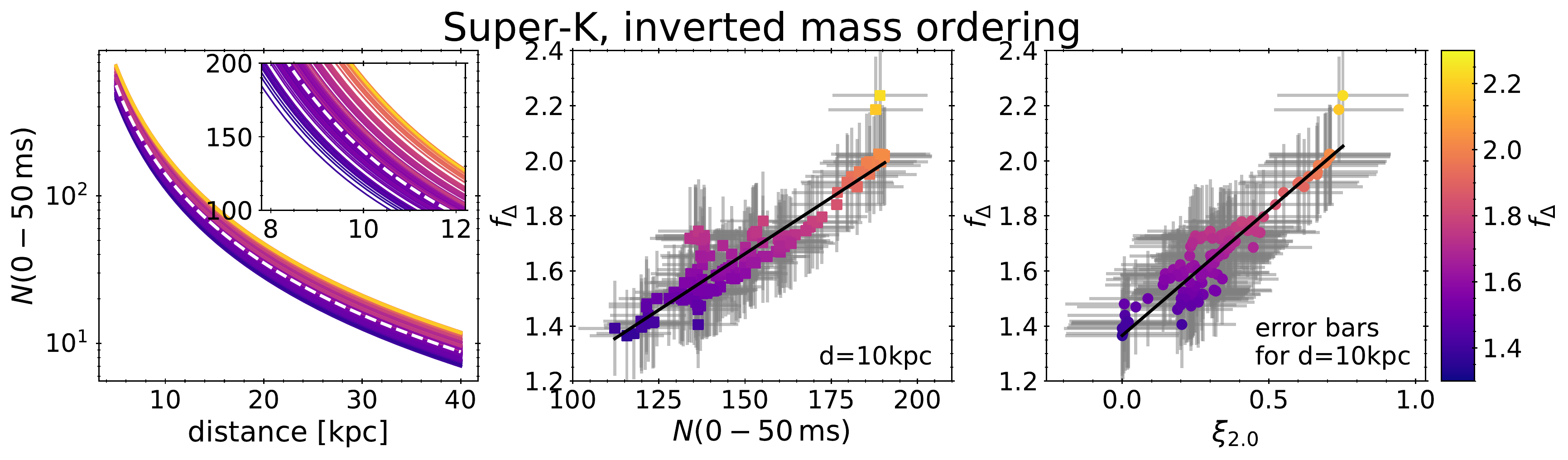}
    \caption{Progenitor dependence of early neutrino signal in the Super-K detector assuming a normal mass ordering (upper panel) and the inverted mass ordering (bottom panel) for the neutrinos. For details, see the Figure 1 caption in the main article.}
    \label{fig:SK}
\end{figure}

\begin{figure}[h]
    \centering
    \includegraphics[width=0.8\textwidth]{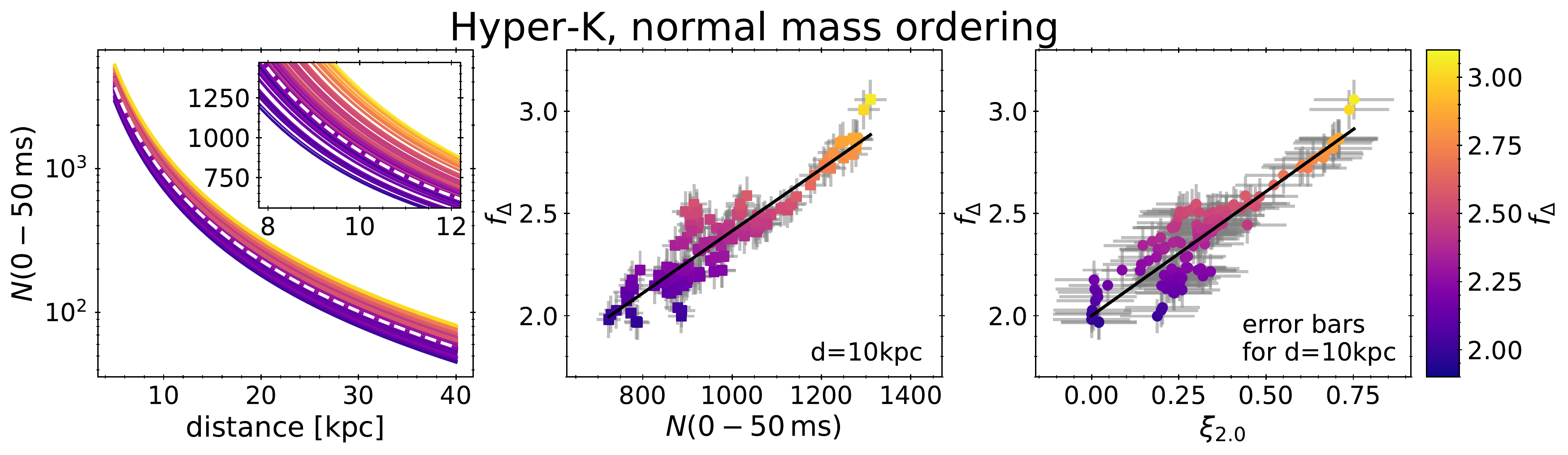}\\
    \includegraphics[width=0.8\textwidth]{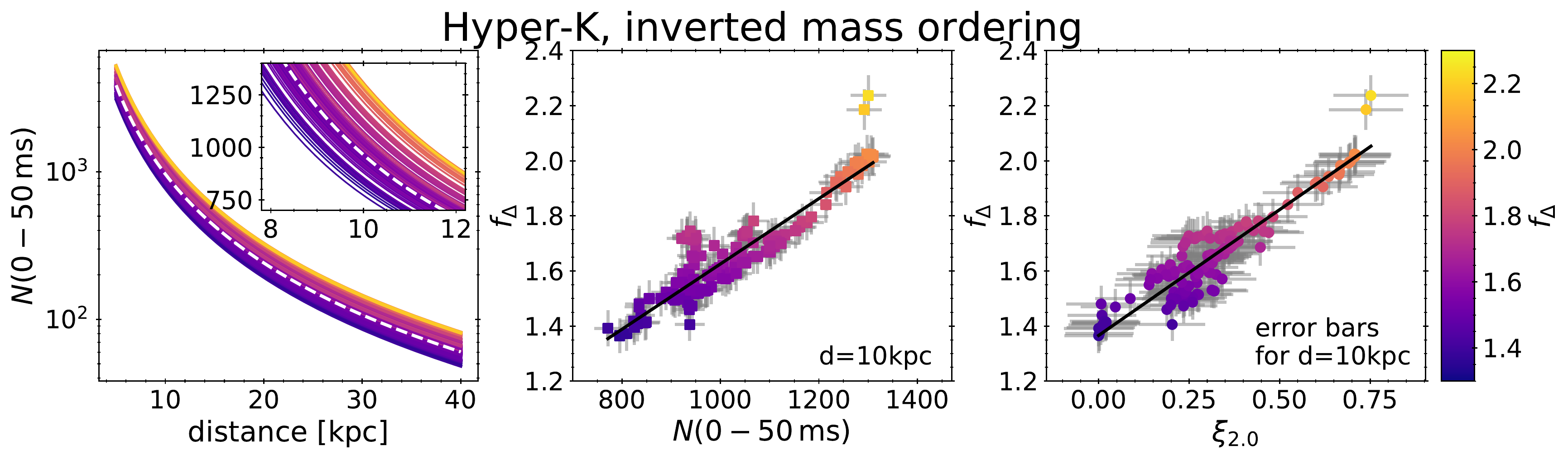}
    \caption{Progenitor dependence of early neutrino signal in the Hyper-K detector assuming a normal mass ordering (upper panel) and the inverted mass ordering (bottom panel) for the neutrinos. For details, see the Figure 1 caption in the main article.}
    \label{fig:HK}
\end{figure}

\begin{figure}[h]
    \centering
    \includegraphics[width=0.8\textwidth]{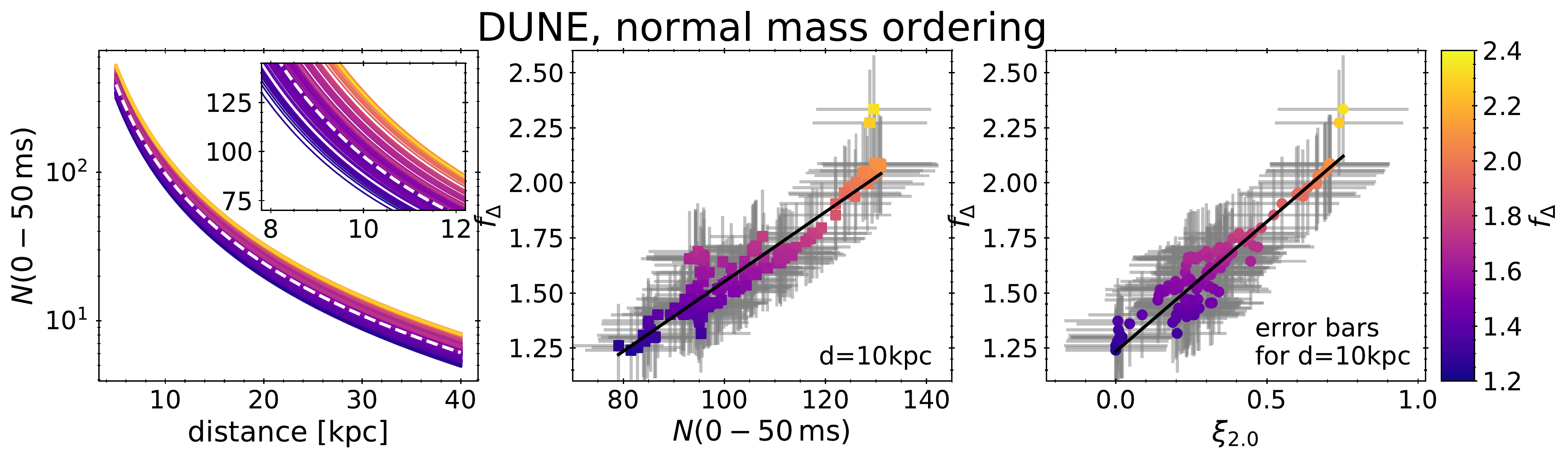}\\
    \includegraphics[width=0.8\textwidth]{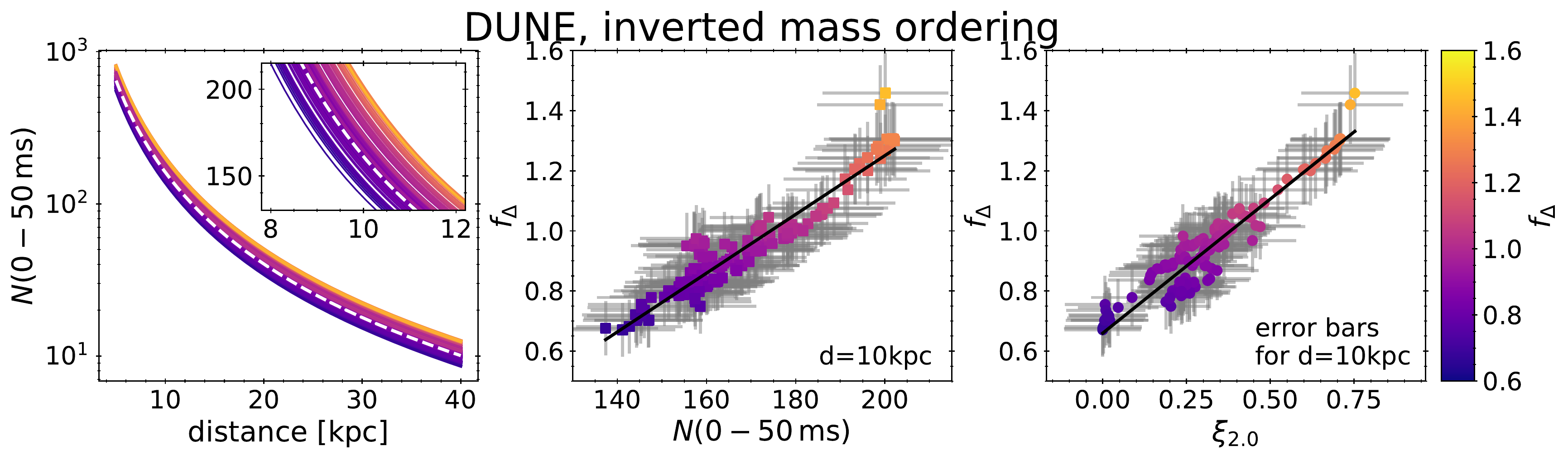}
    \caption{Progenitor dependence of early neutrino signal in the DUNE detector assuming a normal mass ordering (upper panel) and the inverted mass ordering (bottom panel) for the neutrinos. For details, see the Figure 1 caption in the main article.}
    \label{fig:DUNE}
\end{figure}

\begin{figure}[h]
    \centering
    \includegraphics[width=0.8\textwidth]{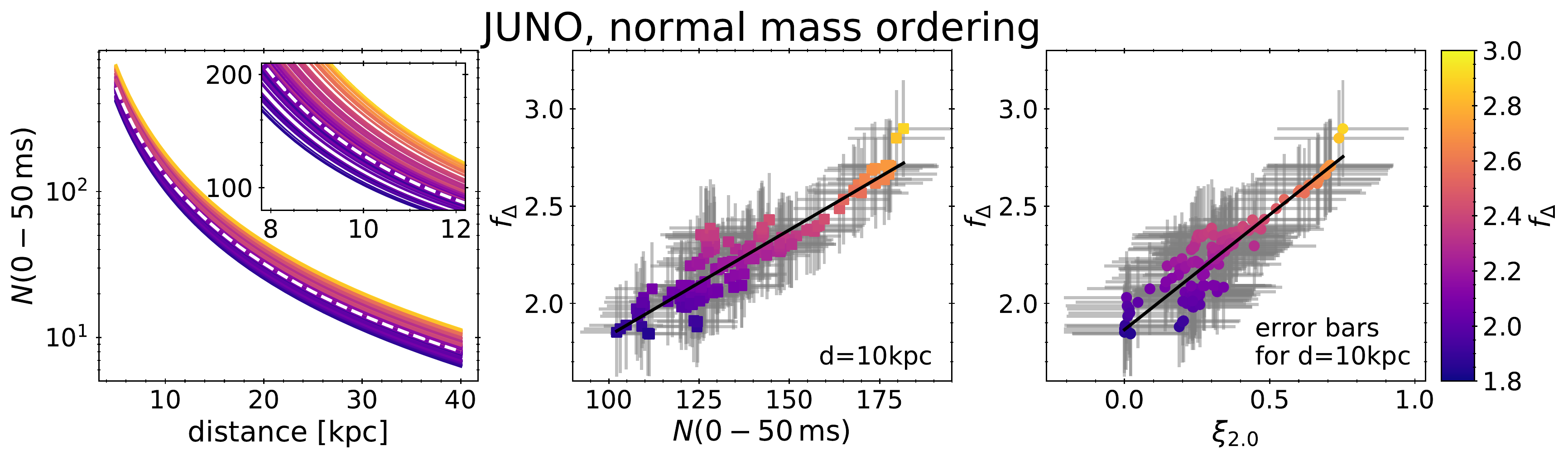}\\
    \includegraphics[width=0.8\textwidth]{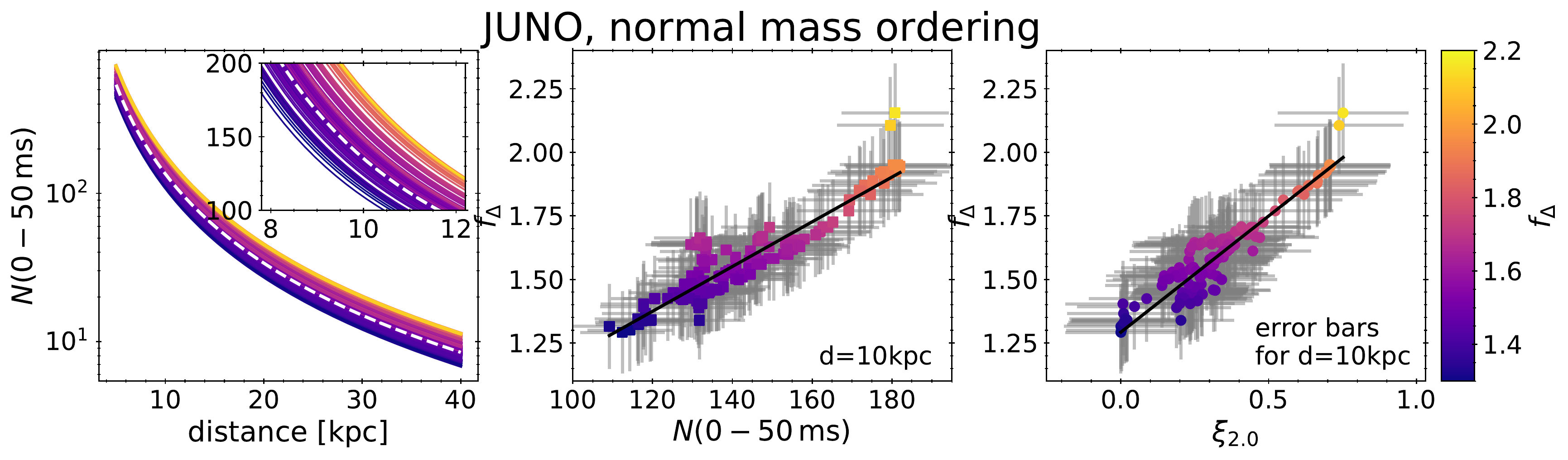}
    \caption{Progenitor dependence of early neutrino signal in the JUNO detector assuming a normal mass ordering (upper panel) and the inverted mass ordering (bottom panel) for the neutrinos. For details, see the Figure 1 caption in the main article.}
    \label{fig:JUNO}
\end{figure}

In the main article we specifically demonstrate our method using the IceCube detector and assuming the normal mass ordering for neutrinos, with the aid of Figure 1.  Here we present the equivalent figures for the inverted mass ordering and the IceCube detector (Fig.~\ref{fig:IC}; for completeness we reproduce the normal mass ordering figure for IceCube as well), for both mass orderings for the Super-K detector (Fig.~\ref{fig:SK}), for the Hyper-K detector (Fig.~\ref{fig:HK}), for the DUNE detector (Fig.~\ref{fig:DUNE}), and for the JUNO detector (Fig.~\ref{fig:JUNO}).

The fits shown in Figs.~\ref{fig:IC}-\ref{fig:JUNO} are given in Table~\ref{tab:fits}. We also include the variance of the models from the fit, which is used as an estimate of the systematic error in our analysis.
\begin{table}[b]
\begin{tabular}{c|ccc|ccc}
Detector, $\nu$ mass ordering & $m_N$ & $b_N$ & $\sigma^\mathrm{sys}_{N,\ b}$ & $m_\xi$ & $b_\xi$ & $\sigma^\mathrm{sys}_{\xi,\ b}$ \\
\hline
\hline
IceCube, NO & 0.000182 & 0.779 & 0.11 & 1.48 & 2.1 & 0.119 \\
IceCube, IO & 0.000125 & 0.342 & 0.0656 & 1.12 & 1.42 & 0.0695 \\
Super-K, NO & 0.0105 & 0.894 & 0.0973 & 1.22 & 2.0 & 0.0996 \\
Super-K, IO & 0.00815 & 0.439 & 0.0529 & 0.914 & 1.37 & 0.0583 \\
Hyper-K, NO & 0.00152 & 0.894 & 0.0973 & 1.22 & 2.0 & 0.0996 \\
Hyper-K, IO & 0.00119 & 0.439 & 0.0529 & 0.914 & 1.37 & 0.0583 \\
DUNE, NO & 0.0158 & -0.0304 & 0.0641 & 1.18 & 1.23 & 0.0696 \\
DUNE, IO & 0.00978 & -0.706 & 0.0411 & 0.893 & 0.66 & 0.05 \\
JUNO, NO & 0.0109 & 0.746 & 0.0909 & 1.18 & 1.87 & 0.0952 \\
JUNO, IO & 0.0088 & 0.319 & 0.0515 & 0.914 & 1.29 & 0.0569 \\
\end{tabular}
\caption{Fit parameters for the relationship between $f_\Delta$ and $N(0-50\,\mathrm{ms})$ and compactness ($\xi_{2.0}$) in the middle and right column, respectively, for the five detectors considered (IceCube, Hyper-K, Super-K, DUNE, and JUNO) and the two neutrino mass orderings (NO: normal mass ordering, IO: inverted mass ordering).  The fits are of the form $f_\Delta = m_{X} X + b_X$.  The variance of the 149 progenitor models to the fit are also given, this is taken as a proxy for the systematic error on the fit.}
\label{tab:fits}
\end{table}

\section{Cumulative Distributions of error on distance and compactness determination}

\begin{figure}[h]
    \centering
    \includegraphics[width=0.5\textwidth]{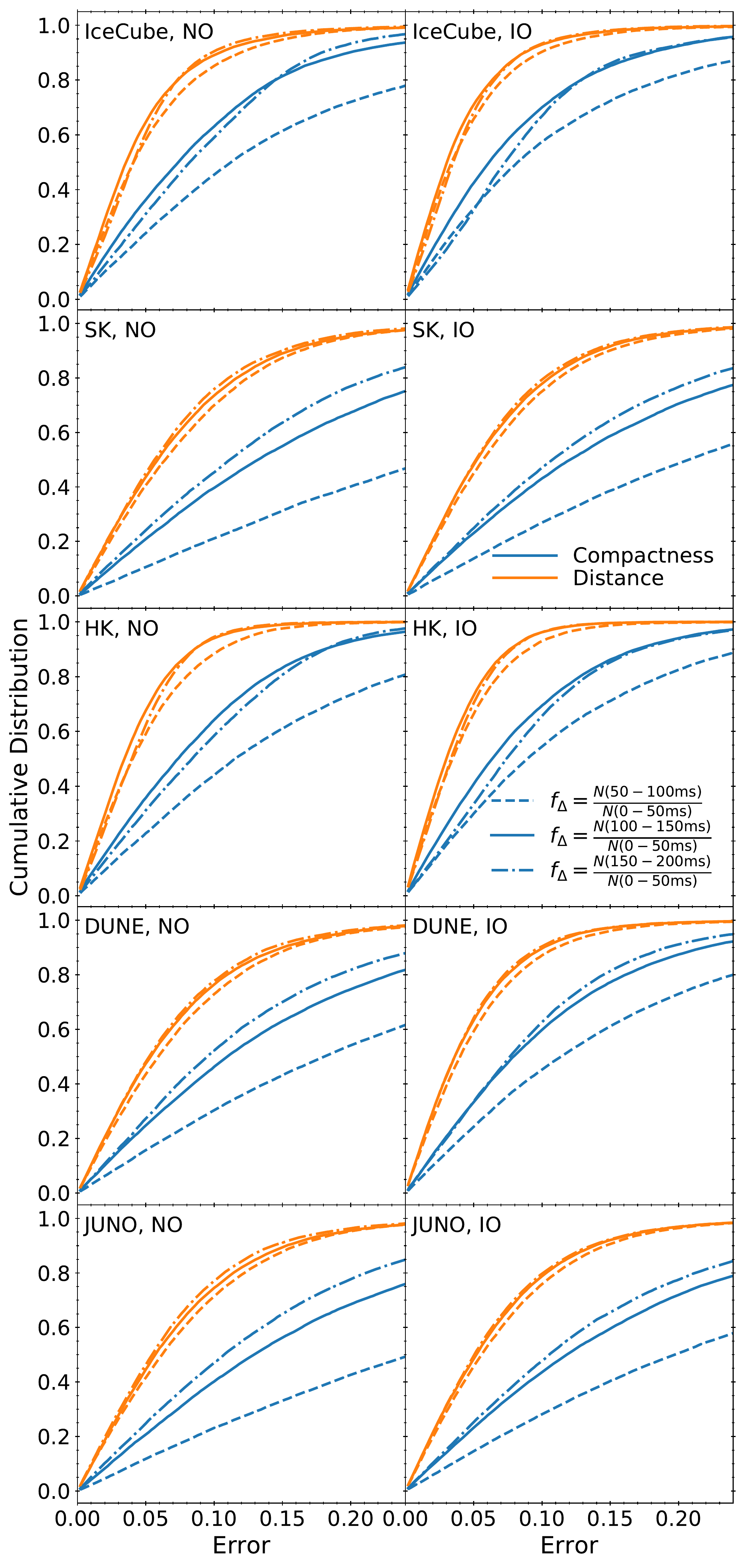}
    \caption{Cumulative distributions of the relative distance error (orange) and the absolute compactness error (blue) marginalized over a galactic population of core-collapse supernovae (i.e. both spatial position \citep{adams:2013} and initial mass function \cite{salpeter:55}).  From the top to the bottom we show the distributions for IceCube, Super-K Hyper-K, DUNE, and JUNO, respectively. The left panels are for the normal neutrino mass ordering while the right panels are for the inverted mass ordering.  The three lines for each detector and mass ordering represent three different defintions of $f_\Delta$, see the text.}
    \label{fig:cumulativedistros}
\end{figure}

In the main article we provide 1$\sigma$ error estimates for the relative error on the distance, and absolution error on the compactness extracted from our mock observations.  For this we marginalized over a galactic spatial distribution \cite{adams:2013} and a Salpeter initial mass function \cite{salpeter:55}.  We show, for each detector and neutrino mass ordering, the full cumulative distribution of this relative error for distance and absolute error on compactness in Fig.~\ref{fig:cumulativedistros}. We show the results for three different definitions of $f_\Delta$ from \cite{horiuchi:2017}.  These three definitions are the ratio of the number of counts in three time windows (50\,ms-100\,ms, 100\,ms-150\,ms, and 150\,ms-200\,ms) to the number of counts in the first 50\,ms.  For the determination of the distance, all three choices are comparable, although with a slight preference for the latest time window, 150\,ms-200\,ms.  For the compactness, the latest time window is generally the best, although for IceCube and Hyper-K the 100\,ms-150\,ms time window is comparable or slightly better.  We settle on $f_\Delta= \frac{N(100-150~\mathrm{ms})}{N(0-50~\mathrm{ms})}$ because the multidimensional dynamics (not included here) will impact the latest time window most significantly. 

\section{Demonstration of distance extraction methods}

\begin{figure}[h]
    \centering
    \includegraphics[width=0.48\textwidth]{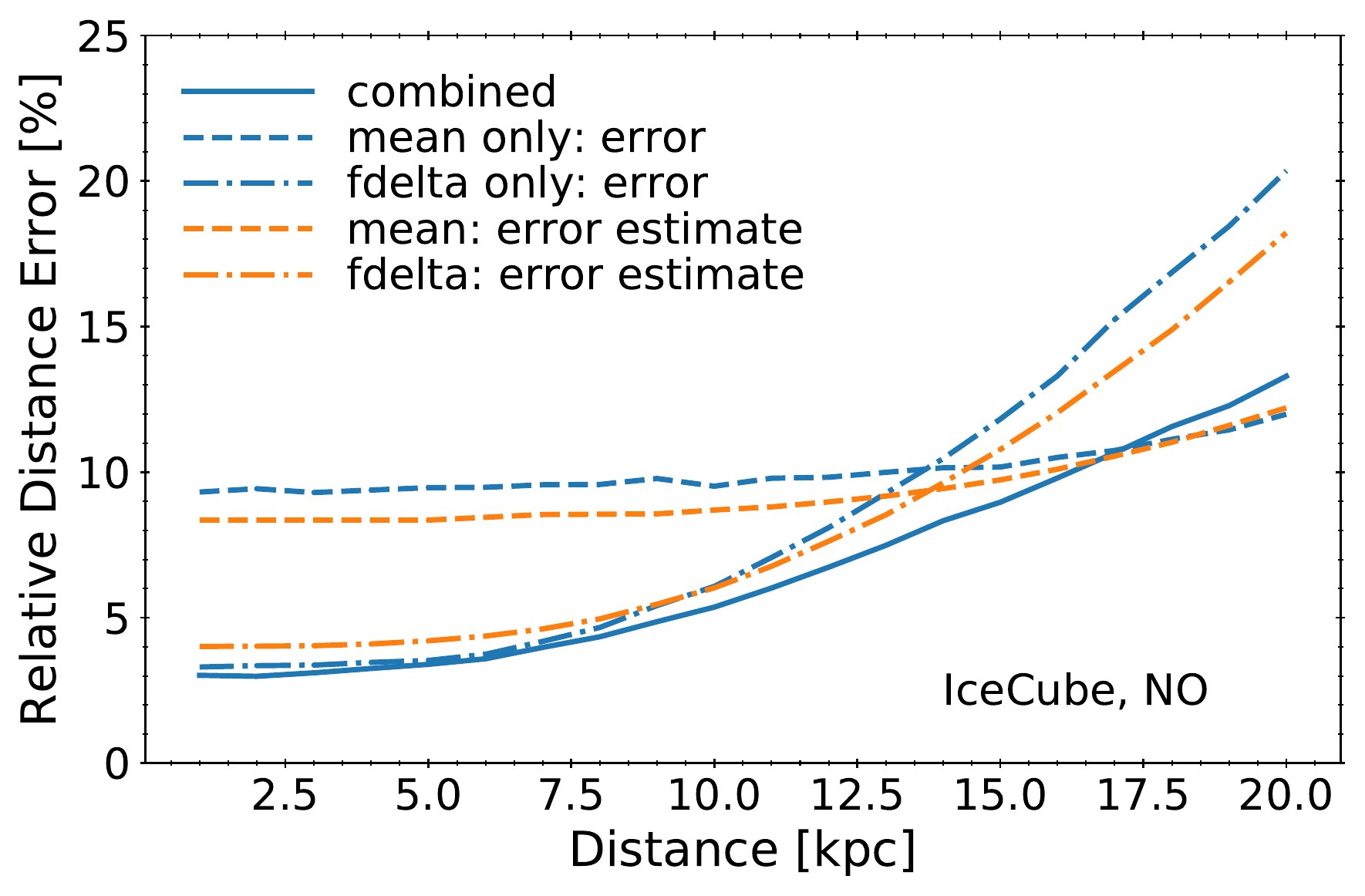}
    \includegraphics[width=0.48\textwidth]{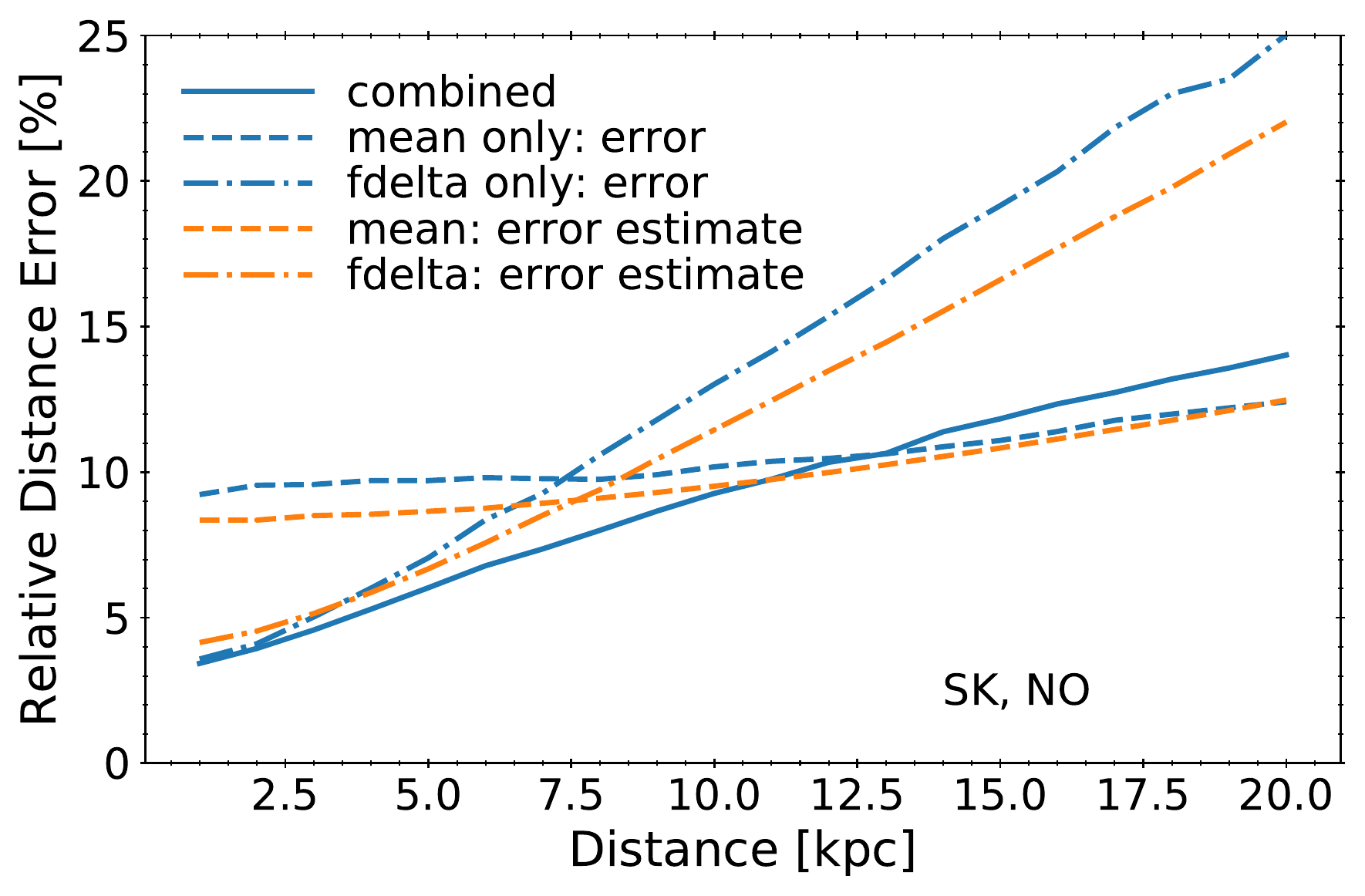}
    \caption{1$\sigma$ relative distance errors for different distance extraction techniques (dashed and dashed-dotted lines) as well as the combination (solid lines).  See the text for details.  We show the results for two detectors: IceCube with normal neutrino mass ordering (left); and Super-K with normal neutrino mass ordering (right).}
    \label{fig:distance_methods}
\end{figure}

We utilize two methods to extract the distance from the early neutrino signal.  One is purely based on the number of counts in the first 50\,ms and compares this observation to the mean progenitor from our model set. The other method first constrains the progenitor model with $f_\Delta$ in order to more precisely determine the expected number of the count.  At close distances (or large detectors) we can constrain the progenitor well enough for the latter method to give better distances estimates.  However, at large distances, or smaller detectors, the error introduced by the progenitor identification is larger than the one introduced by just assuming the mean progenitor.  We explicitly show this in Fig.~\ref{fig:distance_methods} for IceCube and Super-K (both using the normal mass ordering). As a function of distance, the blue dashed and dashed-dotted lines are the 1$\sigma$ relative distance errors for the first (with the mean progenitor model) and the second (via the progenitor constraint with $f_\Delta$) distance-estimate method, respectively.  As argued above, the $f_\Delta$ method performs better at close distances and for larger detectors.  The cross over point is at $\sim$13\,kpc for IceCube and $\sim$7.5\,kpc for Super-K. The orange lines shown in the figures are 1$\sigma$ values of the estimated error which are used to weight the distance estimates from the two methods for each mock observation.  As can be seen, they track the actual errors quite well. The blue solid line is the 1$\sigma$ relative distance error achieved when we combine the two distance-estimate methods.

\end{document}